\documentclass[aps,twocolumn]{revtex4}

\usepackage{graphics}
\usepackage{epsfig}
\usepackage{amsmath}

\newcommand{\Ket}[1]{|#1\rangle}
\newcommand{\Bra}[1]{\langle#1|}

\begin{document}
\title{{\bf Quantum computing with spatially delocalized qubits}}
\author{J.~Mompart$^{a,b}$, K.~Eckert$^{a}$, W. Ertmer$^{c}$, G. Birkl$^{c}$, and M. Lewenstein$%
^{a}$}
\address{$^{a}$Institute of Theoretical Physics, University of Hannover,
Appelstr. 2, D-30167 Hannover, Germany\\
$^b$ Departament de F\'{\i}sica, Universitat Aut\`onoma de Barcelona, 
E-08193 Bellaterra, Spain \\
$^{c}$ Institute of Quantum Optics, University of Hannover, Welfengarten 1,
D-30167 Hannover, Germany}
\date{\today}

\begin{abstract}

We analyze the operation of quantum gates for neutral atoms with qubits that 
are delocalized in space, i.e., the computational basis states are defined by 
the presence of a neutral atom in the ground state of one out of two trapping 
potentials. The implementation of single qubit gates as well as a controlled 
phase gate between two qubits is discussed and explicit calculations are presented 
for rubidium atoms in optical microtraps. Furthermore, we show how multi-qubit
highly entangled states can be created in this scheme. 


\end{abstract}

\maketitle

Optical lattices \cite{olat} and arrays of optical microtraps \cite{PRLBirkl} 
are promising candidates for the implementation
of quantum information processing with neutral atoms.
Many of the requirements for quantum computation with optical microtraps
have been recently demonstrated in 2D arrays of $\sim 80$ traps with $\sim 100$
atoms per trap \cite{PRLBirkl}.
Some remarkable
characteristics of optical microtraps are the possibility to scale, miniaturize and
parallelize the required atom optics devices. 
In addition, they offer two fundamental advantages 
over optical lattices: (i) the possibility of individually addressing single traps 
due to the large separation of the microlenses foci, e.g., $\sim 125 \mu {\rm m}$; and 
(ii) the independent displacement of rows and columns of microtraps and, eventually, 
of single microtraps. Single atoms in dipole traps \cite{singledt}
and the Mott insulator transition with one atom per trap in optical lattices \cite{singleMott} 
have been reported,
and, therefore, 
the achievement of 1D and 2D arrays of optical microtraps 
containing none or one atom per trap in a deterministic way
can be foreseen for the near future. 
We will make use of all these features of optical microtraps to 
propose a novel implementation for quantum information processing.  


In our scheme, each qubit consists of
two traps separated by a distance $2a$ and one single atom. 
Per definition, the detection of the
atom in the ground state of the left trap represents $\left| 0\right\rangle $ 
and in the right trap $\left| 1\right\rangle $, i.e., 
$\left| 0\right\rangle = \left| 0\right\rangle_L$ and
$\left| 1\right\rangle = \left| 0\right\rangle_R$, where 
$\left| 0\right\rangle_{L,R}$ are the vibrational ground states of the left and right trap, 
respectively. Throughout the paper we will call this
implementation the {\it spatially delocalized qubit} (SDQ), since
$\left| \Bra{0} \vec{r} \Ket{0}- \Bra{1} \vec{r} \Ket{1} \right| =2a$
with $\vec{r}$ the position operator.
To implement the SDQ we will assume that we are able to deterministically store
none or one single atom per trap and cool it to the vibrational ground state in 3D.

Single and two-qubit gate operations
will be performed by adiabatically approaching two traps which will be modeled as 
follows: The initial separation of the traps is $2a_{max}$. 
The process of approaching them to the minimum separation
$2a_{min}$ takes a raising time $t_r$. The temporal evolution of the distance $a$ is 
described by the first half of a period of a 
cosine. The two wells remain at the minimum separation for an interaction time $t_i$ and, finally, 
are adiabatically separated to the initial distance.
To simplify the numerical analysis we will assume 
piecewise harmonic trapping potentials as in ref.~\cite{ourQCpaper} and, eventually, 
consider realistic Gaussian potentials as they are present in the experiment \cite{PRLBirkl, BirklOC}.

Single-qubit operations, e.g., a Hadamard gate, are performed by
adiabatically approaching the traps and allowing tunneling to take place. In
order to illustrate this operation, it is convenient to consider the two
lowest energy eigenstates of the double well potential. These two states are
symmetric and antisymmetric, denoted by $\left| S\right\rangle $ and $\left|
A\right\rangle $ respectively, with energies $E_{S,A}(a)=E(a)\mp {1 \over{2}} 
\hbar  {\Omega } (a)$ and $\Omega $ being the splitting frequency. 
In terms of these states, our qubit basis reads: $\left|
0\right\rangle =\frac{1}{\sqrt{2}}\left( \left| S\right\rangle +\left|
A\right\rangle \right) $, and $\left| 1\right\rangle =\frac{1}{\sqrt{2}}%
\left( \left| S\right\rangle -\left| A\right\rangle \right) $. Let us assume
that the atom is initially in the left trap, i.e, $\left| \psi
(t=0)\right\rangle =\left| 0\right\rangle $, then it is straightforward to
check that its time evolution will be given by 
\begin{equation}
\left| \psi (t)\right\rangle =e^{-i\frac{Et}{\hbar }}\left[ \cos \left(
{\Omega \over{2}} t \right) \left| 0\right\rangle 
+i \sin \left( {\Omega \over {2}} t \right) \left| 1\right\rangle \right].
\end{equation}
Thus, the atomic wavefunction oscillates in a Rabi-type fashion between left and right
traps at the flopping frequency ${1 \over{2}} \Omega (a)$. 
Obviously, for large trap
separations states $\left| S\right\rangle $ and $\left| A\right\rangle $
become degenerate in energy, i.e., $\Omega (a)=0$ for $a\rightarrow 
\infty $, and then the atom does not evolve in time (up to a trivial phase).
Therefore, it is possible to realize single-qubit operations via tunneling 
by experimentally controlling the ``Rabi frequency'' through $a_{\min }$, $t_{r}$ and $t_{i}$.

\begin{figure}
%
%
\begin{center}
\medskip
\includegraphics[width=8.0cm]{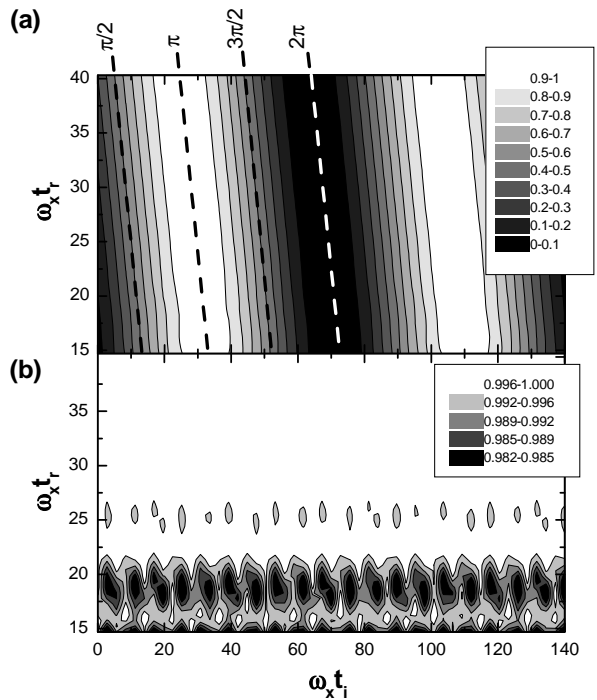}
\end{center}
\label{rabi}
\caption{Single qubit operations for $\alpha a_{max}=5$, $\alpha a_{min}=1.8$ 
and the atom initially in the left trap:
(a) Population of the right trap;
(b) The sum of the ground state population of left plus right traps.
$ \alpha^{-1}=\sqrt{ \hbar / m \omega_x }$ is the position spread of the ground state 
with $m$ the mass of the neutral atom and $\omega_x$ the trapping frequency.
}
\end{figure}

These single-qubit operations are illustrated in Fig.~1 through a numerical
integration of the 1D Schr\"{o}dinger equation in the parameter plane $t_{r}$
versus $t_{i}$ with $\left| 0\right\rangle $ being the initial state. Fig.~1
shows (a) the population of state $\left| 1\right\rangle $, denoted by $\rho
_{1}$, and (b) the total ground state population of left and right traps, i.e., $%
\rho _{0}+\rho _{1}$, after the whole cycle of approaching and separating
the traps. This oscillating population resembles the Rabi-flopping in
the interaction of a laser with a two-level system and, for this reason, 
we have added in Fig.~1 dashed lines indicating the 
$\frac{\pi }{2},\pi ,\frac{3\pi }{2},2\pi $ laser pulse notation conventionally 
used in quantum optics. For small $t_{r}$, non-adiabaticity
results in the population of excited vibrational states, which, as 
shown in Fig.~1 (b), yields $\rho _{0}+\rho _{1}<1$. In what follows, 
we will focus only in the adiabatic regime.

For the two-qubit gate operations we assume that the two qubits
are arranged either (i) in a 1D configuration, i.e., the four traps form a line, or
(ii) side-by-side in a 2D configuration, i.e., the traps form a square, c.f.~Fig.~2(a).
The traps involved are labeled $A_0$, $A_1$ for the first qubit and $B_0$, $B_1$
for the second, and the respective ground states are denoted $\Ket{0}_A$, $\Ket{1}_A$ and
$\Ket{0}_B$, $\Ket{1}_B$. 
A two-qubit gate will be realized via the collisional interaction between bosonic neutral atoms \cite{na2}.
We will consider $^{87}$Rb atoms whose collisional interaction can be 
described by a contact potential of the form 
$U(\vec{r}_1,\vec{r_2})=4\pi a_t\hbar^2 m^{-1}\delta^3(\vec{r}_1-\vec{r_2})$, 
where $a_t=106\,a_0=5.61\,{\rm nm}$ is the $s$-wave scattering length in the spin triplet.
Because the states $\Ket{0}$ and $\Ket{1}$ are localized in different positions,
it is enough to perform a suitable spatiotemporal variation of the potentials in order to pick up a collisional
phase shift e.g.~only if both atoms are in $\Ket{1}$. This is sufficient to implement a two-qubit phase gate,
which transforms product states $\Ket{i}_A\Ket{j}_B$, $i,j\in\{0,1\}$,
into $\exp({i\,\delta_{i1}\delta_{j1}\pi})\Ket{i}_A\Ket{j}_B$ and, supplemented by arbitrary single-qubit gates,
forms a universal set of gates.

For the in-line arrangement, the
change of the potential leading to a phase gate is shown in Fig.~2(b), 
where horizontal and vertical axes denote space and time, respectively.
As the first step a $\pi$ pulse is applied on the second qubit, exchanging
$\Ket{0}_B$ and $\Ket{1}_B$. During this step only single-particle phases arise which can be included
into the definition of the single-particle states. 
If the initial state was $\Ket{0}_A\Ket{0}_B$ then after the $\pi$ pulse, traps $A_1$ and $B_0$ would contain no atom.
For initial states $\Ket{1}_A\Ket{0}_B$ or $\Ket{0}_A\Ket{1}_B$, an atom would be either in $A_1$ or in $B_0$ and
as seen before, we could approach and eventually separate $A_1$ and $B_0$ such that
a $n2\pi$ pulse is applied with $n$ integer. 
In this case initial and final state coincide, except for a single particle phase
$\phi_S$ which again can be included into the definition of $\Ket{1}_A$ or $\Ket{0}_B$.
If we started from $\Ket{1}_A\Ket{1}_B$, then after the first pulse $A_1$ and $B_0$ would both be occupied,
and during the $n2\pi$ pulse the two atoms would collide.
For an adiabatic evolution we can neglect not only the probability to populate
excited vibrational states,
but also to find two atoms in the same trap, since,
due to the collisional interaction, these states are not degenerated with states where
each atom occupies a different trap \cite{ourQCpaper}.
Thus for $a_t\neq0$ initial and final state are the same except for a phase $\phi$ and, in order to realize the 
desired phase gate operation, we need its collisional part $\phi_C=\phi-2\phi_S$  to be an odd multiple of $\pi$.
The fidelity of this operation can be expressed as $F=\rho\cdot(\cos(\phi_C-\pi)+1)/2$, which is plotted
for the adiabatic regime in Fig.~2(c). 
Here $\rho$ is the final probability to find the atom in the same state as it was before 
the $n2\pi$ pulse, neglecting the collisional phase.
To calculate the collisional phase $\phi_C$, we have integrated the two-particle 1D Schr{\"o}dinger equation 
replacing $U(\vec{r}_1,\vec{r}_2)$ by an effective 1D interaction potential under the assumption that no transverse
excitations occur \cite{na2}. Finally, to complete the phase gate operation another $\pi$ pulse is 
applied to the second qubit.

In the case of 2D arrays of traps, as they are typically realized in the experiment \cite{PRLBirkl}, 
the easiest operation is to move complete columns of microtraps. To realize the gate
it is enough to be able to move selectively some columns, with the additional
benefit that the operation is applied to many pairs of qubits in parallel which might 
allow for an easy implementation of error correcting codes.
For the side-by-side arrangement, the initial and final $\pi$ pulse can be omitted
and only the $2\pi$ pulse between traps $A_1$ and $B_1$ is needed. 
Although conceptually much more easier, the implementation in this arrangement demands the ability to move 
single traps instead of columns which makes it experimentally more involved.

\begin{figure}
%
%
\begin{center}
\includegraphics[width=7.5cm]{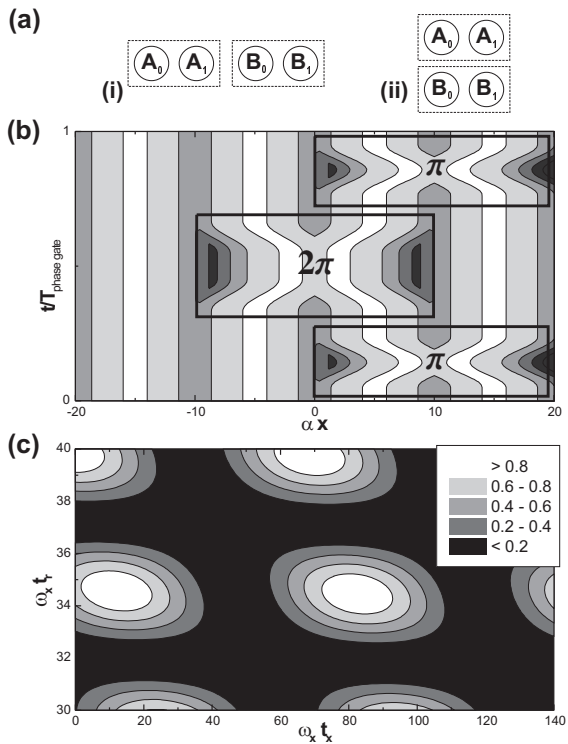}\\
\end{center}
\label{phase gate}
\caption{
Two-qubit phase gate operation. (a) Arrangements of the qubits: 
(i) in-line, and (ii) side-by-side;
(b) Contourplot of the spatiotemporal variation of the trapping potential for arrangement (i). 
The centers of the four traps are white, dark grey means high potential energy;
(c) Fidelity F=$\rho\cdot (\cos(\phi_C-\pi)+1)/2$. Parameters as in Fig.~1 with  
$a_t=106\,a_0$, $\omega_x=\omega_y=2\cdot 10^5\,s^{-1}$, and $\omega_z=1.1\cdot10^6\,s^{-1}$.  
}
\end{figure}
 
The SDQ scheme allows also for the creation of maximally entangled states
in a very straightforward way. Let us consider again four microtraps in the side-by-side arrangement, Fig.~3(a), 
with the two atoms located in the upper-left and lower-right traps, respectively, 
i.e., the initial two particle state is $\Ket{0}_A \Ket{1}_B$. 
During  an approach of the traps, Fig.~3(b), the general two-particle state of the system will be:
\begin{eqnarray}
\Ket{\Psi (t)}&=&
c_{00}\Ket{0}_A \Ket{0}_B + 
c_{01}\Ket{0}_A \Ket{1}_B + \nonumber \\
&&
c_{10}\Ket{1}_A \Ket{0}_B + 
c_{11}\Ket{1}_A \Ket{1}_B + \nonumber \\
&&
\sum_{j=A,B} c_j \Ket{0}_j \Ket{1}_j +
\sum_{i=0,1} \sum_{j=A,B} c_{ij} \Ket{i}_j \Ket{i}_j.
\end{eqnarray}
Thus, the state of the system includes the four states of the computational basis, but also 
double qubit occupation, defined as $\rho_{{\rm \,dq}}$=$\sum_{j=A,B} c_jc_j^*$, 
and double trap occupation,  $\rho_{{\rm \,dt}}$=
$ \sum_{i=0,1} \sum_{j=A,B} c_{ij} c_{ij}^* $.
In order to create maximally entangled states, the approach of the traps must be both, 
adiabatic and symmetric, i.e, $a(t)=b(t)$. Again, adiabaticity means that the population of
excited vibrational states, as well as of double trap occupation states, can be neglected,
c.f.~Fig~3(c).
A symmetric approach results in $\rho_{00} (t)=\rho_{11} (t)=\rho_{{\rm \,dq}}(t)/2$
during the whole process. In particular, these 
populations oscillate at the same frequency and, at the oscillation nodes, 
the state of the system is a combination only of 
$\Ket{0}_A \Ket{1}_B$ and $\Ket{1}_A \Ket{0}_B$. 
Therefore, choosing appropriate parameter values, such as those of Fig.~3, 
it is possible to obtain a maximally entangled state at the end of the process. 
Clearly, it is straightforward to generalize this approach to more 
than two qubits and, consequently, to prepare multiparticle entangled states 
in one single step. 

\begin{figure}
%
%
\begin{center}
\medskip
\includegraphics[width=9.0cm]{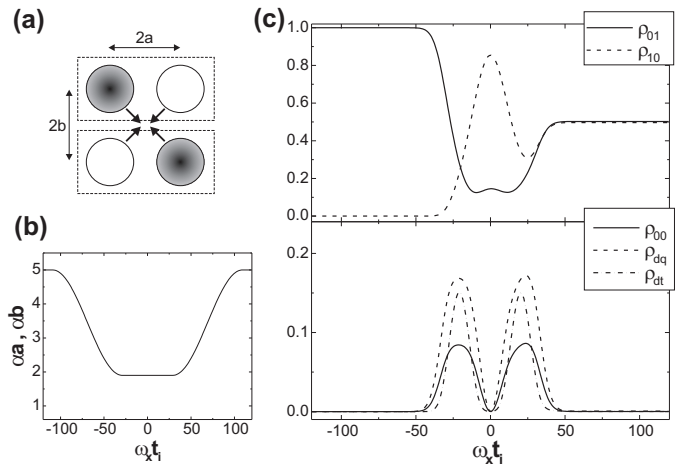}
\end{center}
\caption{
Creation of a maximally entangled two-qubit state with two $^{87}$Rb atoms in a single step:
(a) The initial state is $\Ket{0}_A \Ket{1}_B $, 
with $2a$ and $2b$ being the separation
of the traps in $x$ and $y$ direction, respectively; (b) We adiabatically and 
simultaneously, i.e., $a(t)=b(t)$, approach 
the four traps towards the center of the square; 
(c) Above: Population of states $\Ket{0}_A \Ket{1}_B$ and 
$\Ket{1}_A \Ket{0}_B$; Below: Population of the state $\Ket{0}_A \Ket{0}_B$ (equal to the population of
$\Ket{1}_A \Ket{1}_B$),
double qubit, and double trap occupations.   
The parameters are: $\alpha a_{max}=5$, $\alpha a_{min}=1.9$, 
$\omega_x t_r=80$, $\omega_x t_i =58$, $a_t=106 a_0$, 
$\omega_x = \omega_y=2 \cdot 10^5\,s^{-1}$, and $\omega_z = 1.1 \cdot 10^6\,s^{-1}$.
}
\end{figure}

Although we have assumed harmonic trapping potentials so far, the experimental situation
is described by Gaussian potentials of the form 
$V(x)=-V_0 \exp(-(1/2)\,m\omega_x^2 x^2/V_0)$.
An analysis of the energy eigenvalues and eigenstates of the superposition of two 
such potentials as a function of the trap separation shows that the cosine function 
previously used to adiabatically approach the traps leads to values of $t_r$ larger 
by more than two orders of magnitude compared to harmonic traps.
We have therefore applied the techniques from \cite{optimization} to optimize 
the temporal variation of the trap separation while suppressing the population 
of excited vibrational states.
For $V_0=200\,\hbar\omega_x$, 
Fig.~4(a) shows the result of this optimization for single-qubit operations, 
where the optimization is done with respect to the symmetric ground
and first excited states. We notice that for the minimal distance $2a_{min}$ 
the two traps are no longer separated by a tunneling barrier, but they form a 
single flat trap. Fig.~4(b) shows the population of the right trap, 
$\rho_1$. The error rate due to the excitation of other vibrational states can
be made smaller than  $1\%$ for $w_xt_r>1100$ which is a reduction by one order of
magnitude compared to the non-optimized function $a(t)$.

\begin{figure}
%
%
\begin{center}
\includegraphics[width=8cm]{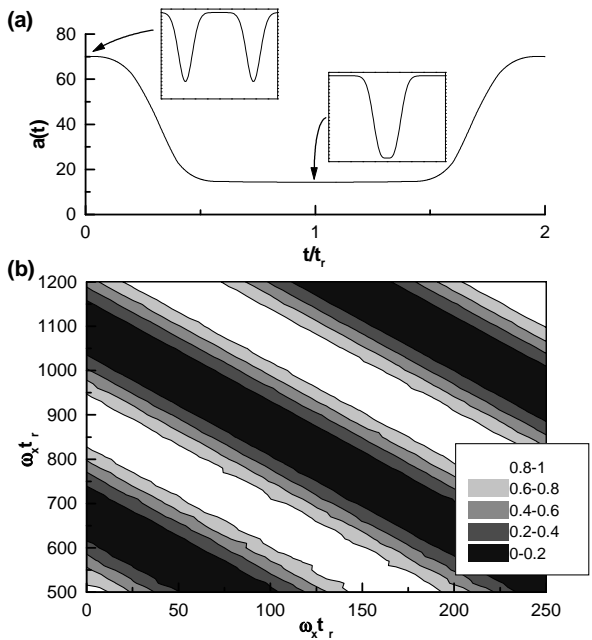}
\end{center}
\caption{
Single-qubit gates for Gaussian trapping potentials for the following parameters:
$\omega_x=6\cdot10^{5}\,s^{-1}$, $\alpha^{-1}=35.4\,\rm{nm}$, 
$V_0=200\,\hbar\omega_x=0.9 \rm{mK}\cdot k_B$,
$\alpha a_{max}=70$, $\alpha a_{min}=14.35$:
(a) The optimized variation of the distance for approaching the traps,
insets show the form of the potential;
(b) Population $\rho_1$ of the right trap showing Rabi-type oscillations.
}
\end{figure}

Typical trapping frequencies for state-of-the-art 2D optical microtraps arrays of $^{87}$Rb atoms are 
$10^5$-$10^6$~s$^{-1}$ in the transverse directions and
$10^4$-$10^5$~s$^{-1}$ along the laser beam direction \cite{PRLBirkl,BirklOC}. 
In order not to populate excited vibrational states and neglect double trap
occupation, the trap displacement has to be adiabatic with respect to the lowest relevant
trapping frequency. 
This adiabaticity condition yields realization times for single and two-qubit operations on the order 
of $1$~ms and $10$~ms, respectively. These values should be compared with the typical 
lifetime of the atoms 
in the microtraps of $\sim 1$~s, and the rate of spontaneous 
scattering of photons from the trapping laser of
$1$-$10$~s$^{-1}$. 
The rate of decoherence for qubits encoded in external states is further reduced by a
factor of 10-100 due the strong localization and almost harmonic trapping potential \cite{echo}.
In addition, sideband cooling can be used to cool the atoms to the ground state 
of each trap in all dimensions. 

As discussed here, the SDQ configuration with neutral atoms 
in optical microtraps presents important practical advantages with respect 
to the use of either internal states \cite{na2,na} 
or vibrational states \cite{vibrational,ourQCpaper} to encode the qubit.
(i) Spontaneous emission leads to decoherence only in a much reduced fashion. 
As long as the microtraps are moved adiabatically, 
atoms remain in the ground state of both the internal and 
external degrees of freedom for typical gate times. (ii) There is no momentum transfer in single or two-qubit
operations as, in general, it is the case when these operations are realized via laser
pulses. This momentum transfer could heat the atoms and, eventually, take them out 
from the microtrap. (iii) There is no need of state dependent interaction to realize the
two-qubit operations. In the SDQ configuration one has that 
$\left| \Bra{0} \vec{r} \Ket{0} - \Bra{1} \vec{r} \Ket{1} \right| \neq 0 $ and, therefore,
we can make use of the fact that all interactions are space dependent to realize the gate.
(iv) The measurement of the state of the system is straightforward. Optical microtraps can be separated
to distances well beyond $\sim 10 \mu {\rm m}$ which allows to detect the population of the trap by
focusing a laser field in one single trap and detecting the fluorescence signal \cite{PRLBirkl}. 
Finally, (v) single and two-qubit gates are realized by the same kind of operation, i.e. 
by approaching the microtraps, which implies a strong simplification in the experimental set-up.   
 
Finally, it is important to note that most of the concepts developed in this Letter can
be also applied to quantum dots with the qubit encoded in two tunnel-split ground 
states and the Coulomb interaction between electrons used to realize the qubit operations \cite{qdots}; 
and to Josephson-junctions based on the charge degree of freedom 
with the Cooper-pairs tunneling coherently through the superconducting junction \cite{Josephson}.

This work is supported by the European Commission through IST projects EQUIP and ACQUIRE,
as well as through a Marie Curie Fellowship
under contract HPMF-CT-2000-00916 (J.M.), 
and by the DFG (Schwerpunktprogramm 'Quanteninformationsverarbeitung' and SFB 407).
We thank D.~Bru{\ss}, R.~Dumke, T.~M\"uther, A.~Sanpera, and M. Volk.


\begin{thebibliography}{0}

\bibitem{olat} P.~S.~Jessen and I.~H.~Deutsch, Adv. At. Mol. Opt. Phys. {\bf 37}, 95 (1996);
I.~Deutsch and P.~S.~Jessen, Phys. Rev. A {\bf 57}, 1972 (1998); 
G.~Grynberg and C.~Robbilliard, Phys. Rep. {\bf 355}, 355 (2001).  

\bibitem{PRLBirkl} R.~Dumke {\it et al.}, Phys. Rev. Lett. {\bf 89}, 097903 (2002). 

\bibitem{singledt} D.~Frese {\it et al.}, Phys. Rev. Lett. {\bf 85}, 3777 (2000); 
N.~Schlosser {et al.}, Nature {\bf 441}, 1024 (2001).

\bibitem{singleMott} M.~Greiner {\it et al.}, Nature (London) {\bf 415}, 39 (2002).

\bibitem{ourQCpaper} K.~Eckert {\it et al.}, to be published in Phys. Rev. A (2002).

\bibitem{BirklOC} G.~Birkl {\it et al.} Optics Comm. {\bf 191}, 67 (2001).

\bibitem{na2}  D.~Jaksch {\it et al.}, Phys. Rev. Lett. {\bf 82}, 1975 (1999);
T.~Calarco {et al.}, Phys. Rev. A {\bf 61}, 022304 (2000).

\bibitem{optimization} W.~H\"ansel {\it et al.}, Phys. Rev. A {\bf 64}, 063607 (2001).

\bibitem{na} G.~K.~Brennen {\it et al.}, Phys. Rev. Lett. {\bf 82}, 1060 (1999);
D.~Jaksch {\it et al.}, Phys. Rev. Lett. {\bf 85}, 2208 (2000);
I.~E.~Protsenko {\it et al.}, Phys. Rev. A {\bf 65}, 052301 (2002).

\bibitem{vibrational} E.~Charron {\it et al.}, Phys. Rev. Lett. {\bf 7}, 077901 (2002).

\bibitem{echo} F.~B.~J.~Buchkremer {\it et al.}, Phys, Rev. Lett. {\bf 85}, 3121 (2000).

\bibitem{qdots} T.~Brandes and T.~Vorrath, to be published in Phys. Rev. B (2002);
F.~Renzoni and T.~Brandes, Phys. Rev. B {\bf 64} 245301 (2001).

\bibitem{Josephson} See Y.~Makhlin {\it et al.}, Rev. Mod. Phys. {\bf 73}, 357 (2001) and references 
therein.

\end{thebibliography}
\end{document}